\documentclass[%
 reprint,
superscriptaddress,
 amsmath,amssymb,
 aps,
pra,
]{revtex4-2}
\usepackage{subfigure} 
\usepackage{tikz}
\usepackage{diagbox}
\usepackage{graphicx}
\usepackage{dcolumn}
\usepackage{bm}
\usepackage{hyperref}
\usepackage[mathlines]{lineno}
\usepackage[linesnumbered,ruled,vlined]{algorithm2e}

\begin{document}

\preprint{APS/123-QED}

\title{Asymmetric protocols for mode pairing quantum key distribution with finite-key analysis}

\author{Zhenhua Li}
\affiliation{China Telecom Research Institute, Beijing 102209, China}
\author{Tianqi Dou}
\affiliation{China Telecom Research Institute, Beijing 102209, China}
\author{Yuheng Xie}
\affiliation{China Telecom Research Institute, Beijing 102209, China}
\author{Weiwen Kong}
\affiliation{China Telecom Research Institute, Beijing 102209, China}
\author{Yang Liu}
\affiliation{China Telecom Research Institute, Beijing 102209, China}
\author{Haiqiang Ma}
\email{hqma@bupt.edu.cn}
\affiliation{School of Science and State Key Laboratory of Information Photonics and Optical Communications, Beijing University of Posts and Telecommunications, Beijing 100876, China}
\author{Jianjun Tang}
\email{tangjj6@chinatelecom.cn}
\affiliation{China Telecom Research Institute, Beijing 102209, China}


\begin{abstract}
The mode pairing quantum key distribution (MP-QKD) protocol has attracted considerable attention for its capability to ensure high secure key rates over long distances without requiring global phase locking. However, ensuring symmetric channels for the MP-QKD protocol is challenging in practical quantum communication networks. Previous studies on the asymmetric MP-QKD protocol have relied on ideal decoy state assumptions and infinite-key analysis, which are unattainable for real-world deployment. In this paper, we conduct a security analysis of asymmetric MP-QKD protocol with the finite-key analysis, where we discard the previously impractical assumptions made in the decoy-state method. Combined with statistical fluctuation analysis, we globally optimized the 12 independent parameters in the asymmetric MP-QKD protocol by employing our modified particle swarm optimization. The simulation results demonstrate that our work can achieve significantly enhanced secure key rates and transmission distances compared to the original strategy with adding extra attenuation. We further investigate the relationship between the intensities and probabilities of signal, decoy, and vacuum states with transmission distance, facilitating its more efficient deployment in future quantum networks.
\end{abstract}

\maketitle

\section{Introduction}
Classical encryption methods rely on computational complexity, which exposes them to the risk of being compromised by future advances in computility. In contrast, quantum key distribution (QKD) guarantees the absolute security of information transmission by leveraging the fundamental principles of quantum mechanics \cite{ekert1991quantum,lo1999unconditional,shor2000simple}. 
Since the inception of the first QKD protocol, namely the BB84 protocol \cite{BEN84}, it has ignited a surge of research interest in the field of QKD. In QKD systems, single-photon serves as the carriers of quantum keys, which cannot be amplified and are easily scattered or absorbed by the transmission channel. For the transmittance channel $\eta$, the secure key rate (SKR) cannot exceed  the PLOB (Pirandola, Laurenza, Ottaviani, and Banchi) bound $R\le-\log_2\left(1-\eta\right)$ \cite{pirandola2017fundamental}, which represents the exact SKR limit for point-to-point QKD systems without quantum repeaters.

Twin-field (TF) QKD \cite{lucamarini2018overcoming} utilizes single-photon interference for key generation, surpassing the PLOB bound, while also retaining immunity to detector-side attacks.  Nevertheless, in order to maintain the coherence between distant quantum states, TF-QKD requires the deployment of global phase locking technology\cite{pittaluga2021600,liu2023experimental}, leading to increased complexity and additional resource consumption within QKD systems. Fortunately, the recently proposed mode pairing (MP) QKD protocol \cite{zeng2022mode,xie2022breaking} eliminates the requirement for global phase locking while also surpassing the repeaterless bound. Given its exceptional performance in laboratory \cite{zhu2023experimental,zhou2023experimental} and field environments \cite{zhu2024field,li2024field}, the MP-QKD protocol has established itself as a robust candidate for the future deployment of quantum communication networks.

In scenarios involving geographical disparities (e.g., in star-shaped quantum communication networks \cite{tang2016measurement}), or when the communicating parties (Alice and Bob) are situated in moving free spaces (e.g., drones, ships, satellites, etc.), challenges are posed in achieving symmetric communication with the central node, Charlie. 
Introducing additional losses to compensate for channel asymmetry is a solution, but it results in suboptimal SKR \cite{rubenok2013real}. Previous analyses of the asymmetric MP-QKD protocol \cite{lu2024asymmetric} required ideal decoy state assumptions and infinite-key, which are overly idealized and impractical. 
Thus, it is crucial to account for the finite-key size effects in the asymmetric MP-QKD protocol, while considering practical decoy state deployment.

In this paper, we evaluate the performance of MP-QKD protocol with asymmetric channels. Through employing the practical decoy state method and the universally composable framework, we analyzed the security of asymmetric MP-QKD protocol with finite-key size. By employing our modified particle swarm optimization (PSO) algorithm, we globally optimized the 12 independent parameters of the decoy state MP-QKD protocol. Our modified PSO algorithm is more suitable for asymmetric MP-QKD, as it does not rely on the specific form or gradient information of the SKR function, demonstrating robust global search capabilities. Numerical simulation results show that when the distance difference between Alice-Charlie and Bob-Charlie is 50 km and 100 km, the SKRs with the asymmetric intensity strategy show a significant improvement compared to the original protocol, though it remains lower than that of the symmetric channel. We further present the relationship curves between signal and decoy state intensities (probabilities) and channel loss, and provide an analysis to explain the observed differences. Finally, we examine the effect of the maximum pairing interval on the performance of asymmetric MP-QKD protocol.

\section{Finite-key analysis for MP-QKD with asymmetric channels}
Since the key size in practical implementations is not infinite, it is crucial to develop a framework that addresses the finite-key size effect, which has not been extensively analyzed in the previous MP-QKD protocol. At the end of the asymmetric MP-QKD protocol, Alice and Bob each share a key string, denoted as $\mathbb{S}$ and $\mathbb{S}'$, respectively. According to the universally composable framework\cite{muller2009composability}, if these key strings satisfy both $\epsilon_{\rm{cor}}$-correct and $\epsilon_{\rm{sec}}$-secret, they are considered secure keys. $\epsilon_{\rm{cor}}$-correct refers to $\Pr \left( {\mathbb{S} \ne  \mathbb{S}'} \right) \le \epsilon_{\rm{cor}}$. And $\epsilon_{\rm{sec}}$-secret requires satisfaction of the condition
\begin{equation}
    \frac{1}{2}{\left\| {{\rho _{AE}} - {U_A} \otimes {\rho _E}} \right\|_1} \le \epsilon_{\rm{sec}},
\end{equation}
where ${\left\|  \cdot  \right\|_1}$ denotes the trace norm, $\rho _{AE}$ is the  density operator of the system of Alice and Eve, $U_A$ is the uniform mixture  of all possible values of the bit string $\mathbb{S}$, and $\rho _E$ is density operator of Eve's system. In this way, the protocol with finite-key can be regarded as $\epsilon$-secure, where $\epsilon = \epsilon_{\rm{cor}} + \epsilon_{\rm{sec}}$. In addition, we fix the security bound to $\epsilon = 10 ^{-10}$.

 
 
In the following finite-key  analysis of the asymmetric MP-QKD protocol, we assume the distances from Alice and Bob to Charlie are denoted as $L_A$ and $L_B$, where $L_A \le L_B$ and $\Delta L = L_{B}-L_{A}$. Here, we focus solely on the three-intensity decoy state asymmetric MP-QKD protocol, as methods such as multi-intensity decoy state\cite{chau2020security} or the double-scanning method\cite{jiang2021higher} fail to enhance the SKR of MP-QKD protocol. 


In the $i$-th round, Alice prepares the coherent state $\left| {{e^{i\theta _a^i}}\sqrt {k_a^i} } \right\rangle$, where intensity $k_a^i$ is randomly selected from $\left\{ \mu_a, \nu_a, o_a \right\}$ with probabilities $\left\{ p_{\mu_a}, p_{\nu_a}, p_{o_a} \right\}$. And modulated phase $\theta _a^i$ is  randomly chosen from $\left\{ {0,\frac{{2\pi }}{\Delta},\frac{{4\pi }}{\Delta},...,\frac{{2\pi \left( {\Delta - 1} \right)}}{\Delta}} \right\}$, where $\Delta$ is typically set to 16. In parallel, Bob randomly selects $k_b^i$ and $\theta _b^i$ to prepare the coherent state $\left| {{e^{i\theta _b^i}}\sqrt {k_b^i} } \right\rangle$ with probabilities $\left\{ p_{\mu_b}, p_{\nu_b}, p_{o_b} \right\}$. It is important to note that in asymmetric channels, except for the case where $o_{a}=o_{b}=0$, the intensity and probability settings of Alice and Bob are unequal. Therefore, full parameter optimization is required to achieve the maximum SKR. After interference measurement, Charlie announces the measurement outcomes ${D_L},{D_R} \in \left\{0,1\right\}$ (0 represents ``no click", while 1 represents ``click"). And ${D_L} \oplus {D_R} = 1$ is considered as an effective detection. After $N$ rounds, Alice and Bob pair each click with its immediate next neighbor within a maximum pairing interval $l$ to form a successful pairing. 

According to the intensities of the paired $i$-th and $j$-th rounds ($j \le i + l$), Alice (Bob) labels the ``basis" as illustrated in Table \ref{basis_assignment}.

\begin{table}[h]
\caption{%
Basis assignment depending on intensities.
}
\begin{ruledtabular}
\begin{tabular}{cccc}
\diagbox{$k_{a(b)}^j$}{$k_{a(b)}^i$} &
$\mu_{a(b)}$ &
$\nu_{a(b)}$ &
$o_{a(b)}$ \\
\colrule
$\mu_{a(b)}$ & $X$-basis & `discard' & $Z$-basis \\
$\nu_{a(b)}$ & `discard' & $X$-basis & $Z$-basis \\
$o_{a(b)}$ & $Z$-basis & $Z$-basis & `$0$'-basis \\
\end{tabular}
\end{ruledtabular}
\label{basis_assignment}
\end{table}

\begin{table}[h]
\caption{%
The pair assignment.
}
\begin{ruledtabular}
\begin{tabular}{cccc}
\diagbox{Bob}{Alice} &
$Z$-basis &
$X$-basis &
`0'-basis \\
\colrule
$Z$-basis & $Z$-pair & `discard' & $Z$-pair \\
$X$-basis & `discard' & $X$-pair & $X$-pair \\
`0'-basis & $Z$-pair & $X$-pair & `$0$'-pair \\
\end{tabular}
\end{ruledtabular}
\label{pair_assignment}
\end{table}

 Subsequently, Alice and Bob announce the basis, and the sum of the intensity pairs $\left( {{k_a},{k_b}} \right) = \left( {k_a^i + k_a^j,k_b^i + k_b^j} \right)$ for each pairing. Subsequently, they perform the pair assignment based on Table \ref{pair_assignment}. 

For each $Z$-pair on location $i$, $j$, Alice
(Bob) extracts a bit 0 when $k_a^i \ne k_a^j = {o_a}$ ($k_b^j \ne k_b^i = {o_b}$), and a bit 1 when $k_a^j \ne k_a^i = {o_a}$ ($k_b^i \ne k_b^j = {o_b}$). For each $X$-pair on location $i$, $j$, Alice (Bob) extracts a bit from the relative phase $\left[ {{{\left( {\theta _{a\left( b \right)}^i - \theta _{a\left( b \right)}^j} \right)} \mathord{\left/
 {\vphantom {{\left( {\theta _{a\left( b \right)}^i - \theta _{a\left( b \right)}^j} \right)} \pi }} \right.
 \kern-\nulldelimiterspace} \pi }} \right]\bmod 2$ and announces the alignment angle ${\theta _{a\left( b \right)}} = \left( {\theta _{a\left( b \right)}^i - \theta _{a\left( b \right)}^j} \right)\bmod \pi $. Alice and Bob only retain the results with $\left| {{\theta _a} - {\theta _b}} \right| \le \Delta$ or $\left| {{\theta _a} - {\theta _b}} \right| \ge \pi - \Delta$. In particular, when $\left| {{\theta _a} - {\theta _b}} \right| \ge \pi - \Delta$, Bob flips the bit. Thus, Alice (Bob) can then obtain the sifted key strings $\mathbb{Z}$ ($\mathbb{Z}'$) and $\mathbb{X}$ ($\mathbb{X}'$), derived respectively from $Z$-pair and $X$-pair. According to the decoy state method, $\mathbb{Z}$ ($\mathbb{Z}'$) is employed for generating secure keys, while $\mathbb{X}$ ($\mathbb{X}'$) is used to estimate the phase-error rate.

In finite-key scenarios, sifted keys may contain some errors. Here, error correction and privacy amplification are required to ensure both $\epsilon_{\rm{cor}}$-correct and $\epsilon_{\rm{sec}}$-secret of the keys. Alice sends $\lambda_{\rm{EC}}$ bits to Bob for performing key reconciliation, through which Bob computes an estimate $\hat{\mathbb{Z}}'$, of $\mathbb{Z}'$. Alice computes a hash of ${\mathbb{Z}}$ of length ${\log _2}\frac{2}{{{\varepsilon _{{\rm{cor}}}}}}$ with a random universal$_2$ hash function\cite{carter1977universal}, which she sends to Bob together with the hash. If hash($\hat{\mathbb{Z}}'$) = hash(${\mathbb{Z}}$), this guarantees the $\epsilon_{\rm{cor}}$-correct of the keys; otherwise, the protocol aborts. 
Based on the min-entropy\cite{tomamichel2012tight,tomamichel2011uncertainty}, a random universal$_2$ hash function is used to extract an $\epsilon_{\rm{sec}}$-secret key of length $\mathbf{L}$ from $\mathbb{Z}$, where 
\begin{equation}\label{secc}
    {\epsilon_{\sec }} = 2\varepsilon  + \frac{1}{2}\sqrt {{2^{\mathbf{L} - H_{\min }^\epsilon \left( {\mathbb{Z}|\mathbb{E}'} \right)}}}.
\end{equation}
Here, $\mathbb{E}'$ summarizes all information Eve learned about $\mathbb{Z}$ during the protocol. By solving Eq. (\ref{secc}), we can obtain the secret key length after privacy amplification as
\begin{equation}\label{PAA}
    \mathbf{L}\le H_{\min }^\epsilon \left( {\mathbb{Z}|\mathbb{E}'} \right) - 2\log_{2}\frac{1}{\epsilon_{\rm{sec}}}.
\end{equation}
Based on a chain rule inequality\cite{renner2008security}, the smooth min-entropy $H_{\min }^\epsilon \left( {\mathbb{Z}|\mathbb{E}'} \right)$ can be represented as
\begin{equation}\label{chain}
    H_{\min }^\epsilon \left( {\mathbb{Z}|\mathbb{E}'} \right) \ge H_{\min }^\epsilon \left( {\mathbb{Z}|\mathbb{E}} \right) - {\lambda _{EC}} - {\log _2}\frac{2}{{{\epsilon _{{\rm{cor}}}}}},
\end{equation}
where $\mathbb{E}$ is the information before error correction. Considering the smooth entropies from the uncertainty relation\cite{tomamichel2012tight,curty2014finite}, we further find that
\begin{equation}\label{RST}
    H_{\min }^\epsilon \left( {\mathbb{Z}|\mathbb{E}} \right)\le M_{11}^Z\left[ {1 - h\left( {{e}_{11}^{Z,{\rm{ph}}}} \right)} \right],
\end{equation}
where $h\left( x \right) =  - x{\log _2}x - \left( {1 - x} \right){\log _2}\left( {1 - x} \right)$,  the number of single-photon pair events in the $Z$-basis $M_{11}^Z$ and the phase-error rate associated with the single-photon pair events in the $Z$-basis ${{e}_{11}^{Z,{\rm{ph}}}}$ can be determined using the decoy state method, which will be explained in detail later. By combining Eq. (\ref{PAA}), (\ref{chain}), and (\ref{RST}), the final secure key length is
\begin{equation}\label{RRRR}
    \mathbf{L}\le M_{11}^Z\left[ {1 - h\left( {e_{11}^{Z,{\rm{ph}}}} \right)} \right] - {\lambda _{{\rm{EC}}}} - {\log _2}\frac{2}{{{\varepsilon _{{\rm{cor}}}}}} - 2{\log _2}\frac{1}{{{\varepsilon _{\sec }}}},
\end{equation}
where ${\lambda _{{\rm{EC}}}} = f{M_{\left( {\mu ,\mu } \right)}}h\left( {{E_{\left( {\mu ,\mu } \right)}}} \right)$, $f$ is the error correction efficiency, $M_{\left( {\mu ,\mu } \right)}$ represents the number of pairs ${\left( {\mu ,\mu } \right)}$ for $Z$-basis, and $E_{\left( {\mu ,\mu } \right)}$ denotes the corresponding bit error rate. As long as the final key length satisfies Eq. (\ref{RRRR}), the asymmetric MP-QKD protocol is $\epsilon$-secure. 
The finite-key security analysis based on the entropy uncertainty relation method only requires consideration of the statistical fluctuations in observed quantities, without the need to account for additional information leakage. Here, the Chernoff bound\cite{yin2020tight} is employed to calculate the statistical fluctuations. Given an observed quantity $\chi$, the upper $\overline{\chi}$ and lower $\underline{\chi}$ bounds of the expected value is given by,
\begin{equation}
\begin{aligned}
     \overline{\chi}=& \chi  + \beta  + \sqrt {2\beta \chi  + {\beta ^2}}, \\
\underline{\chi}= &\max \left\{ {\chi  - \frac{\beta }{2} - \sqrt {2\beta \chi  + \frac{{{\beta ^2}}}{4}} ,0} \right\},
\end{aligned}
\end{equation}
where $\beta  = \ln \frac{1}{\epsilon_{\rm{CB}}}$. Here, we set the failure probability ${\epsilon_{\rm{CB}}}$ to $10^{-10}$.

\section{Simulation and discussion}
Since the number of single-photon pair events $M_{11}^Z$ in the $Z$-basis  and the phase-error rate associated with the single-photon pair events ${{e}_{11}^{Z,{\rm{ph}}}}$ in the $Z$-basis  cannot be directly observed, through the decoy state method\cite{wang2005beating,lo2005decoy,xu2013practical}, we can discover the yield of $Z$-pair single photon pulse pairs
\begin{equation}
    y_{11}^Z = \frac{{{F^L} - {F^U}}}{{a_1^{{\nu _a}}a_1^{{\mu _a}}\left( {b_1^{{\nu _b}}b_2^{{\mu _b}} - b_1^{{\mu _b}}b_2^{{\nu _b}}} \right)}},
\end{equation}
where
\begin{widetext}
    \begin{equation}
    \begin{aligned}
{F^L} =& \frac{{a_1^{{\mu _a}}b_2^{{\mu _b}}}}{{N_{\left( {{\nu _a},{\nu _b}} \right)}^Z}}\underline{n}_{\left( {{\nu _a},{\nu _b}} \right)}^Z + \frac{{a_1^{{\nu _a}}b_2^{{\nu _b}}a_0^{{\mu _a}}}}{{N_{\left( {{o_a},{\mu _b}} \right)}^Z}}\underline{n}_{\left( {{o_a},{\mu _b}} \right)}^Z+ \frac{{a_1^{{\nu _a}}b_2^{{\nu _b}}b_0^{{\mu _b}}}}{{N_{\left( {{\mu _a},{o_b}} \right)}^Z}}\underline{n}_{\left( {{\mu _a},{o_b}} \right)}^Z + \frac{{a_1^{{\mu _a}}b_2^{{\mu _b}}a_0^{{\nu _a}}b_0^{{\nu _b}} - a_1^{{\nu _a}}b_2^{{\nu _b}}a_0^{{\mu _a}}b_0^{{\mu _b}}}}{{N_{\left( {{o_a},{o_b}} \right)}^Z}}\underline{n}_{\left( {{o_a},{o_b}} \right)}^Z,\\
{F^U} =& \frac{{a_1^{{\nu _a}}b_2^{{\nu _b}}}}{{N_{\left( {{\mu _a},{\mu _b}} \right)}^Z}}\overline{n}_{\left( {{\mu _a},{\mu _b}} \right)}^Z + \frac{{a_1^{{\mu _a}}b_2^{{\mu _b}}a_0^{{\nu _a}}}}{{N_{\left( {{o_a},{\nu _b}} \right)}^Z}}\overline{n}_{\left( {{o_a},{\nu _b}} \right)}^Z + \frac{{a_1^{{\mu _a}}b_2^{{\mu _b}}b_0^{{\nu _b}}}}{{N_{\left( {{\nu _a},{o_b}} \right)}^Z}}\overline{n}_{\left( {{\nu _a},{o_b}} \right)}^Z.
    \end{aligned}
\end{equation}
\end{widetext}
Here, $a_m^{{k_a}} = \frac{{{{\left( {{k_a}} \right)}^m}{e^{ - {k_a}}}}}{{m!}},b_m^{{k_b}} = \frac{{{{\left( {{k_b}} \right)}^m}{e^{ - {k_b}}}}}{{m!}}$, $\overline{n}_{\left( {{k_a},{k_b}} \right)}^Z$ and $\underline{n}_{\left( {{k_a},{k_b}} \right)}^Z$ are the number of the effective detection for the intensity pairs ${\left( {{k_a},{k_b}} \right)}$ after the statistical fluctuations in the $Z$-pairs, $N_{\left( {{k_a},{k_b}} \right)}^Z$ is the expected number of pairs with  ${\left( {{k_a},{k_b}} \right)}$. For simplicity, here we only use the $Z$-basis pairs with intensities ${\left( {{\mu_a},{\mu_b}} \right)}$ for key generation. Thus, $M_{11}^Z = N_{\left( {{\mu _a},{\mu _b}} \right)}^Z{\mu _a}{\mu _b}{e^{ - {\mu _a} - {\mu _b}}}y_{11}^Z$.

The single-photon bit error rate of $X$-pair can be expressed as
\begin{equation}
    e_{11}^{X,{\rm{bit}}} = \frac{{{T^U} - {T^L}}}{{a_1^{2{\nu _a}}b_1^{2{\nu _b}}y_{11}^Z}},
\end{equation}
where 
\begin{equation}
    \begin{aligned}
{T^U} &= \frac{1}{{N_{\left( {2{\nu _a},2{\nu _b}} \right)}^X}} {\overline{t}_{\left( {2{\nu _a},2{\nu _b}} \right)}^X}   + \frac{{a_0^{2{\nu _a}}b_0^{2{\nu _b}}}}{{N_{\left( {2{o_a},2{o_b}} \right)}^X}} {\overline{t}_{\left( {2{o_a},2{o_b}} \right)}^X}  ,\\
{T^L} &= \frac{{a_0^{2{\nu _a}}}}{{N_{\left( {{0_a},2{\nu _b}} \right)}^X}} {\underline{t}_{\left( {{0_a},2{\nu _b}} \right)}^X}   + \frac{{b_0^{2{\nu _b}}}}{{N_{\left( {2{\nu _a},{o_b}} \right)}^X}} {\underline{t}_{\left( {2{\nu _a},{o_b}} \right)}^X},
    \end{aligned}
\end{equation}
and $\overline{t}_{\left( {{k_a},{k_b}} \right)}^X$ and $\underline{t}_{\left( {{k_a},{k_b}} \right)}^X$ are the error effective detection for the ${\left( {{k_a},{k_b}} \right)}$ after the statistical fluctuations in the $X$-pairs. Through a random-sampling theory without replacement\cite{fung2010practical,lim2014concise,chau2018decoy}, $e_{11}^{Z,\rm{ph}}$ can be written as
\begin{equation}
    e_{11}^{Z,\rm ph} \le e_{11}^{X,\rm bit} + \Gamma\left(\xi_{ee}, e_{11}^{X,\rm bit}, M_{11}^{X}, M_{11}^Z\right),
\end{equation}
where
\begin{equation}
    \Gamma \left( {a,b,c,d} \right) = \sqrt {\frac{{\left( {c + d} \right)\left( {1 - b} \right)b}}{{cd}}\ln \left( {\frac{{c + d}}{{2\pi cd\left( {1 - b} \right)b{a^2}}}} \right)}.
\end{equation}
$\xi_{ee}$ is the failure probability of random sampling without replacement, which we set to $10^{-10}$. $M_{11}^{X}$ represent the number of single-photon pair events in the $X$-basis. After calculating $M_{11}^{Z}$ and $e_{11}^{Z,\rm ph}$, the secure key length for the asymmetric MP-QKD protocol under finite key analysis can be obtained by applying Eq. (\ref{RRRR}). The SKR is defined as $R = 2\mathbf{L}/N$, where $N$ is the number of total pluses.

Here, we can consider the SKR as a function of the source parameters
\begin{equation}
    R\left({\vec g}\right) = R\left( {{\mu _a},{\nu _a},{o_a},{p_{{\mu _a}}},{p_{{\nu _a}}},{p_{{o_a}}},{\mu _b},{\nu _b},{o_b},{p_{{\mu _b}}},{p_{{\nu _b}}},{p_{{o_b}}}} \right).
\end{equation}
Given the uncertainty of the convex form of function $R$, we employ the  modified PSO algorithm (detail in Appendix  \ref{mpso}) for global optimization of the 12 parameters rather than the local search algorithm (LSA)\cite{wang2019asymmetric,wang2020simple,xu2014protocol}. The LSA is highly sensitive to the selection of the initial point, as a randomly chosen initial point can often result in an invalid or infeasible outcome. The modified PSO algorithm can optimize non-smooth and non-convex functions to search for the optimal $\vec g$ that maximizes $R$. It is especially effective for asymmetric MP-QKD because it does not depend on the specific form or gradient information of the SKR function, showing robust global search abilities.


\begin{table}[!htp]
\caption{%
Parameters used in numerical simulation. $p_{d}$ and $\eta_d$ represent the dark counting rate per pulse and detection efficiency of SPD, respectively. $\alpha$ is the transmission fiber loss. $f$ is the error correction efficiency.  $e_{d}^{X}$ and $e_{d}^{Z}$ are the misalignment-error of the $X$-pair and $Z$-pair, respectively. $N$ is the number of total pulses.
}
\begin{ruledtabular}
\begin{tabular}{ccccccc}
$p_{d}$ & $\eta_d$ & $\alpha$ & $f$ & $e_{d}^{X}$ & $e_{d}^{Z}$ & $N$ \\
\colrule
$10^{-8}$ & $75\%$ & 0.2 dB/km & 1.1 & 0.1 & $10^{-6}$ & $10^{13}$ \\
\end{tabular}
\end{ruledtabular}
\label{simulation}
\end{table}

In the simulation, we employed an asymmetric intensity strategy to address the asymmetric channel: by deploying the PSO algorithm to optimize $\vec{g}$ for compensating the channel asymmetry and achieving the optimal SKR. The simulation formulas for the observed values in asymmetric MP-QKD are provided in Appendix \ref{siumlation}.
Furthermore, we also compare this approach with the original strategy of adding extra attenuation. It is important to note that in the simulation, we consider Alice's and Bob's source configurations as independent, satisfying
\begin{gather}\label{condition1}
0 = {o_{a\left( b \right)}} < {\nu _{a\left( b \right)}} < {\mu _{a\left( b \right)}} < 1,\notag\\
{p_{{o_{a\left( b \right)}}}} + {p_{{\nu _{a\left( b \right)}}}} + {p_{{\mu _{a\left( b \right)}}}} = 1,\\
0 \le {p_{{\mu _{a\left( b \right)}}}},{p_{{o_{a\left( b \right)}}}},{p_{{\nu _{a\left( b \right)}}}} \le 1 \notag
\end{gather}
without imposing constraints such as ${\eta _a}{\mu _a} \approx {\eta _b}{\mu _b}$ (or ${\eta _a}{v_a} \approx {\eta _b}{v_b}$), where $\eta_a$ and $\eta_b$ are the channel transmittance of Alice and Bob. The formula ${\eta _a}{\mu _a} \approx {\eta _b}{\mu _b}$ (or ${\eta _a}{v_a} \approx {\eta _b}{v_b}$) is a rule of thumb on the ratio of intensities between Alice's and Bob's light \cite{xu2010experimental}. The parameters employed for the numerical simulations are detailed in Table \ref{simulation}.

\begin{figure}[!htp]
    \centering
    \subfigure[The optimized SKR for $\Delta L = 50$ km.]{%
        \includegraphics[width=0.49\textwidth]{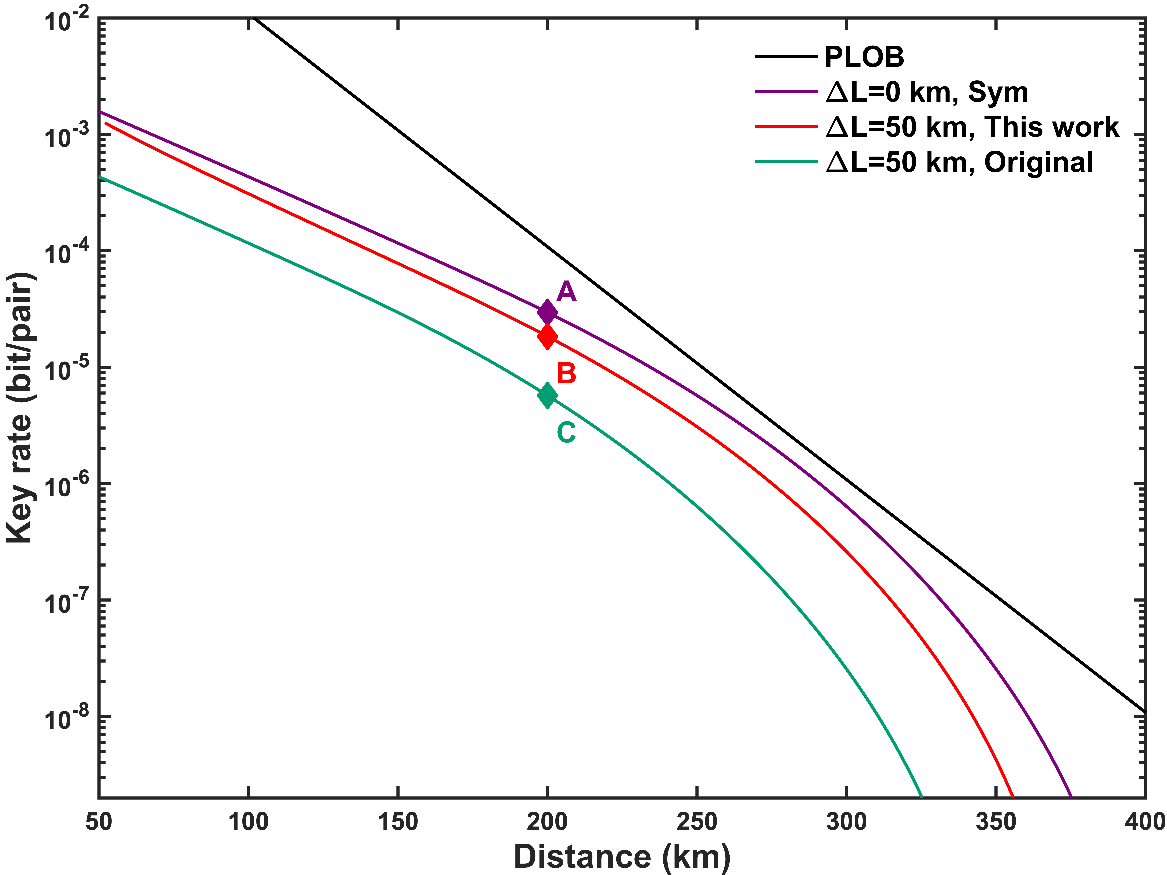}%
        \label{fig:L50}%
    }\hfill
    \subfigure[The optimized SKR for $\Delta L = 100$ km.]{%
        \includegraphics[width=0.49\textwidth]{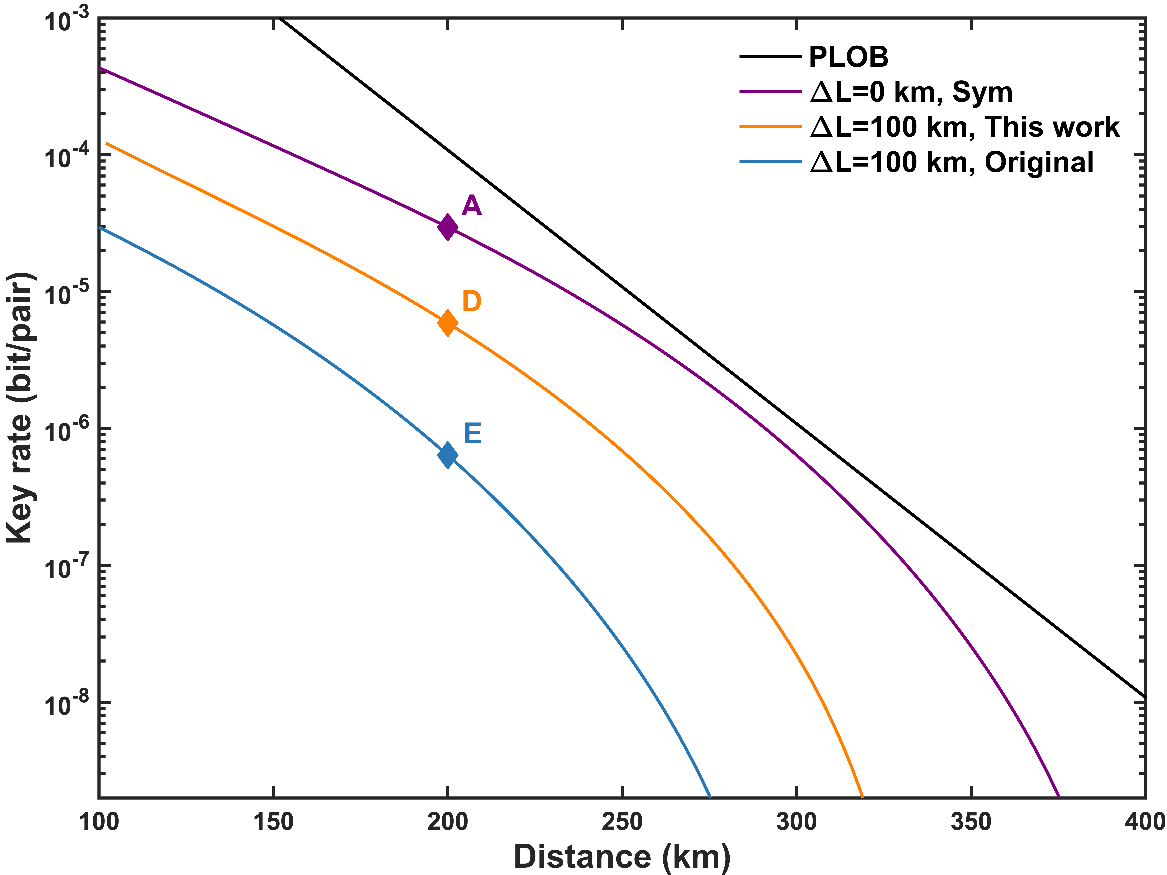}%
        \label{fig:L100}%
    }
    \caption{The optimized SKR versus the total distance ($L_A + L_B$) between Alice and Bob. The maximum pairing interval is fixed to $2000$. The SKR is calculated for two cases, i.e., (a) a 50 km length difference between Alice and Bob, and (b) a 100 km length difference between Alice and Bob. Label 'Sym' represents the symmetric case; Label 'This work' refers to the asymmetric intensity method in this paper; Label 'Original' refers to the original processing method, with added extra attenuation directly. The details of point A, B, C, D, E are given in Table \ref{detail}.}
    \label{len50100}
\end{figure}

\definecolor{AAAA}{rgb}{0.5, 0, 0.5}  
\definecolor{BBBB}{rgb}{1, 0, 0}  
\definecolor{CCCC}{rgb}{0, 0.620, 0.451}  
\definecolor{DDDD}{rgb}{1, 0.5, 0}  
\definecolor{EEEE}{rgb}{0.156, 0.470, 0.710}  

\begin{table*}[!t]
\footnotesize
    \centering
    \caption{Examples of optimal parameters in Fig. \ref{len50100}. The numerical values in the table presented here are rounded to three significant figures.}
    \label{detail}
    \begin{ruledtabular}
        \begin{tabular}{ccccccccccccccc}
            Point & $L_{A}+L_{B}$ & $\Delta L$ & Strategy & $\mu_a$ & $\nu_a$ & $\mu_b$ & $\nu_b$ & $p_{\mu_a}$ & $p_{\nu_a}$ & $p_{o_a}$ & $p_{\mu_b}$ & $p_{\nu_b}$ & $p_{o_b}$ & $R$ (bit/pulse) \\
            \colrule
            \begin{tikzpicture}[scale=0.09]
                \fill[AAAA] (0,1.5) -- (1,3) -- (2,1.5) -- (1,0) -- cycle;
            \end{tikzpicture} A & 
            200 km & 0 km & Sym & 0.424 & 0.0213 & 0.424 & 0.0213 & 0.254 & 0.180 & 0.566 & 0.254 & 0.180 & 0.566 & $2.95\times10^{-5}$ \\
            \begin{tikzpicture}[scale=0.09]
                \fill[BBBB] (0,1.5) -- (1,3) -- (2,1.5) -- (1,0) -- cycle;
            \end{tikzpicture} B & 
            200 km & 50 km & This work & 0.216 & 0.00449 & 0.621 & 0.0376 & 0.170 & 0.229 & 0.601 & 0.305 & 0.192 & 0.503 & $1.84\times10^{-5}$ \\
            \begin{tikzpicture}[scale=0.09]
                \fill[CCCC] (0,1.5) -- (1,3) -- (2,1.5) -- (1,0) -- cycle;
            \end{tikzpicture} C & 
            200 km & 50 km & Original & 0.492 & 0.0258 & 0.492 & 0.0258 & 0.271 & 0.220 & 0.509 & 0.271 & 0.220 & 0.509 & $5.71\times10^{-6}$ \\
            \begin{tikzpicture}[scale=0.09]
                \fill[DDDD] (0,1.5) -- (1,3) -- (2,1.5) -- (1,0) -- cycle;
            \end{tikzpicture} D & 
            200 km & 100 km & This work & 0.107 & 0.000624 & 0.718 & 0.0549 & 0.0902 & 0.309 & 0.6008 & 0.327 & 0.230 & 0.443 & $5.89\times10^{-6}$ \\
            \begin{tikzpicture}[scale=0.09]
                \fill[EEEE] (0,1.5) -- (1,3) -- (2,1.5) -- (1,0) -- cycle;
            \end{tikzpicture} E & 
            200 km & 100 km & Original & 0.560 & 0.0321 & 0.560 & 0.0321 & 0.278 & 0.281 & 0.441 & 0.278 & 0.281 & 0.441 & $6.37\times10^{-7}$ \\
        \end{tabular}
    \end{ruledtabular}
\end{table*}

We plot the optimized SKR for the asymmetric scenario in Fig. \ref{len50100} and the detailed optimization results at a distance of 200 km in Table \ref{detail}.  From Fig. \ref{len50100}, it is evident that the SKR of the asymmetric intensities strategy consistently outperforms that of the original adding extra attenuation, for both $\Delta L = 50$ and $\Delta L = 100$. However, regardless of the strategy employed, the SKR does not achieve a level comparable to that of the symmetric channel case. The SKR for $\Delta L = 100$ experiences a significant decrease compared to $\Delta L = 50$, indicating that an increase in channel asymmetry leads to a reduction in the SKR. As evident from the Table \ref{detail}, the asymmetric intensities strategy exhibits a substantial increase of approximately an order of magnitude compared to the adding extra attenuation strategy when $L_{a}+L_{b}=200$ km. All data in the Table \ref{detail} are optimized using the modified PSO algorithm. Furthermore, from points A, B, and D, it can be observed that Alice and Bob adjusted their intensities (probabilities) to compensate for the channel asymmetry.
As $\Delta L$ increases, to achieve the optimal Hong-Ou-Mandel (HOM) interference effect, $\mu_a$ and $\nu_a$  continuously decrease, while $\mu_b$ and $\nu_b$  increase.


\begin{figure}[!htp]
    \centering
    \includegraphics[width=0.49\textwidth]{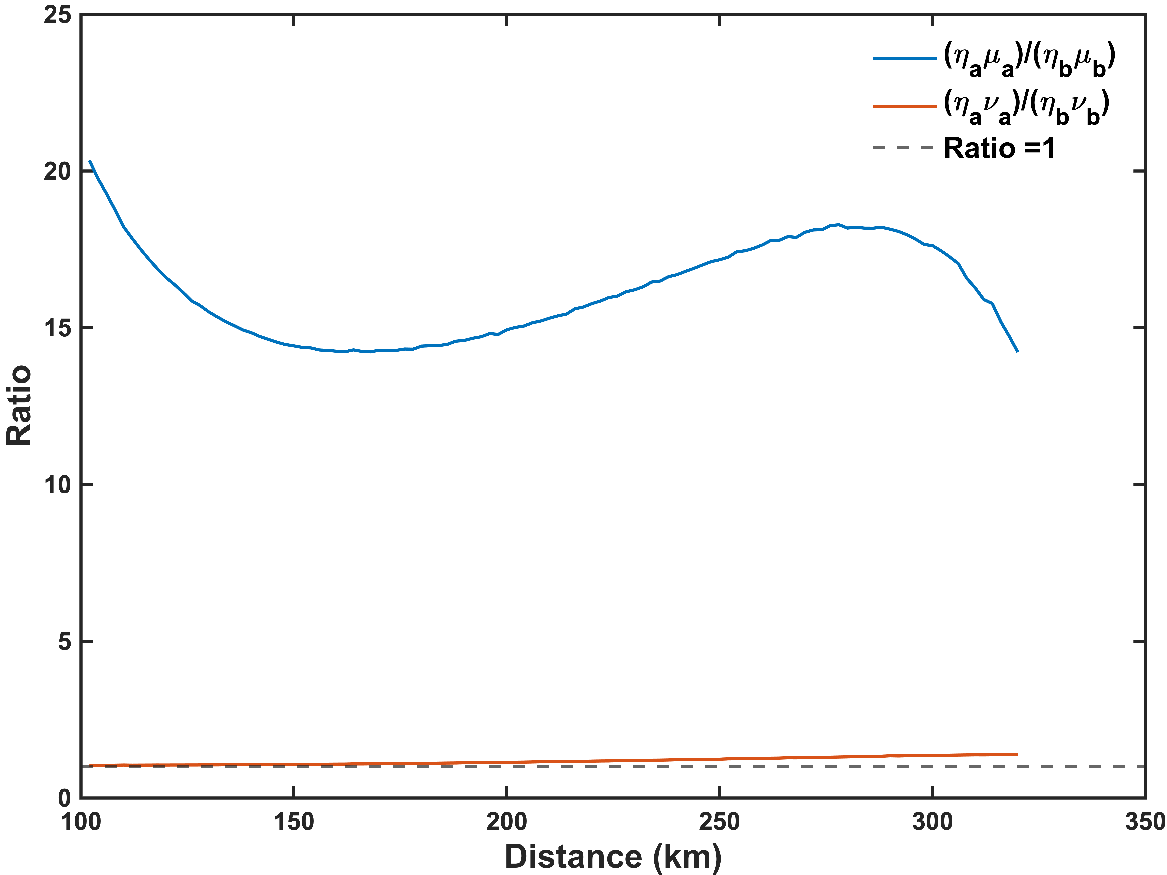}
    \caption{The optimized ratios ${\textstyle{{{\eta _a}{\mu _a}} \over {{\eta _b}{\mu _b}}}}$ and ${\textstyle{{{\eta _a}{\nu _a}} \over {{\eta _b}{\nu _b}}}}$ are plotted versus the total distance ($L_A + L_B$) between Alice and Bob, where $\Delta L=100$ km and the maximum pairing interval $l=2000$.}
    \label{logg}%
\end{figure}

\begin{figure}[!htp]
    \centering
    \includegraphics[width=0.49\textwidth]{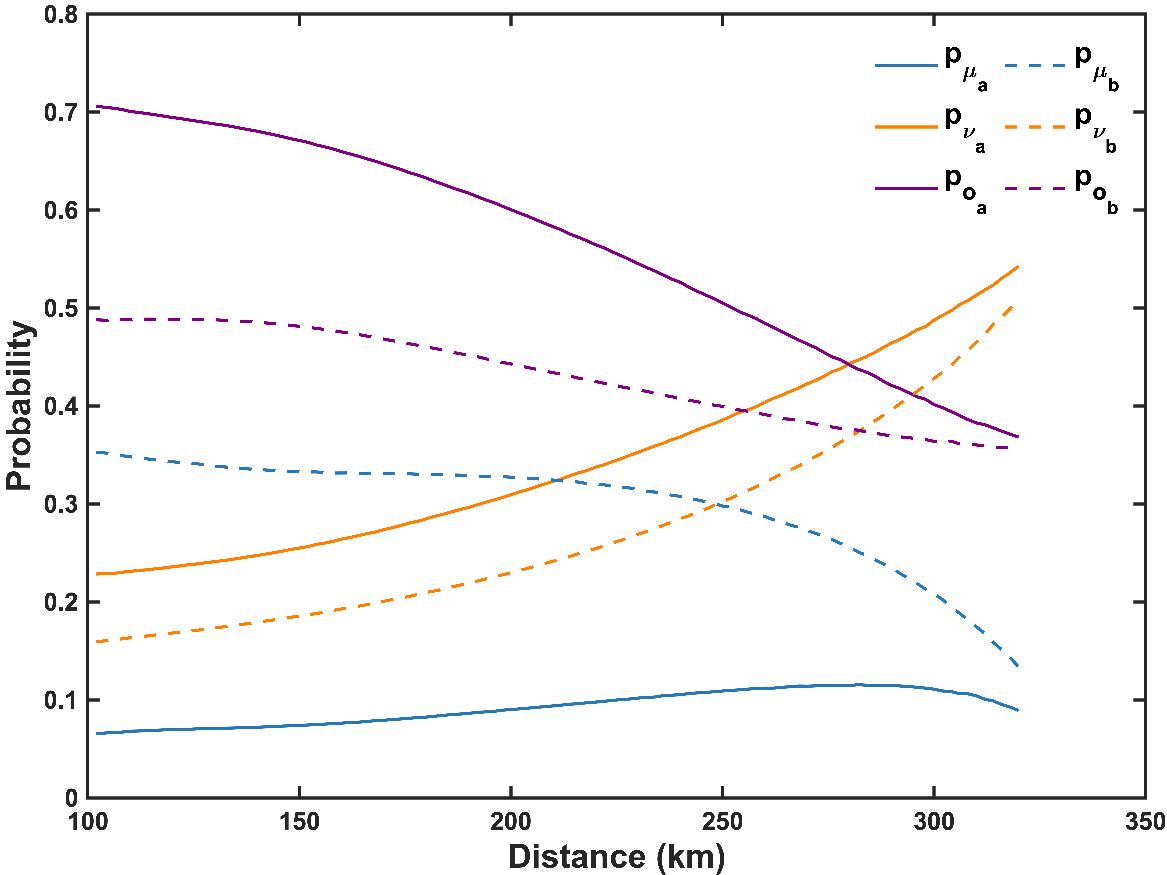}
    \caption{The optimized probabilities of the signal, decoy, and vacuum states with an asymmetric intensity strategy are plotted versus the total distance ($L_A + L_B$) between Alice and Bob, where $\Delta L=100$ km and the maximum pairing interval $l=2000$.}
    \label{probb}%
\end{figure}

In asymmetric intensity strategy for MP-QKD, a primary concern is the optimal selection of weak coherent pulse intensities by Alice and Bob. While ${\eta _a}{\mu _a} \approx {\eta _b}{\mu _b}$ (or ${\eta _a}{v_a} \approx {\eta _b}{v_b}$) may seem natural, we do not impose this constraint in the modified PSO algorithm. Figure \ref{logg} depicts the variation of ${\textstyle{{{\eta _a}{\mu _a}} \over {{\eta _b}{\mu _b}}}}$ and ${\textstyle{{{\eta _a}{\nu _a}} \over {{\eta _b}{\nu _b}}}}$ for $\Delta L=100$ km. It is evident that the magnitude of ${\textstyle{{{\eta _a}{\nu _a}} \over {{\eta _b}{\nu _b}}}}$ remains predominantly near 1, although some deviations occur as the distance increases.  The decoy state $\nu_{a}$ ($\nu_{b}$) are primarily employed for estimating the phase error rate. The relationship ${\eta _a}{v_a} \approx {\eta _b}{v_b}$ helps maintain a balance in the photon intensities reaching Charlie, ensuring good HOM visibility and low error rates. However, this relationship is derived under the assumptions of the infinite-key size, an infinite number of decoy states, and the neglect of dark counts. Therefore, discarding the ideal assumption mentioned above, slight deviations in $\eta_{a} \nu_{a}$ and $\eta_{b} \nu_{b}$ are reasonable for the scenario depicted in Fig. \ref{logg}. Additionally, from Fig. \ref{logg}, it can be observed that the value of ${\textstyle{{{\eta _a}{\mu _a}} \over {{\eta _b}{\mu _b}}}}$ is significantly distant from 1, largely deviating from the condition ${\eta _a}{\mu _a} \approx {\eta _b}{\mu _b}$. The reason for this phenomenon is that the signal state is primarily used for key generation, where the intensity $\mu_{a}$ ($\mu_{b}$) not only impacts the quantum bit error rate but also influences the probability of sending single photons and error correction. The optimal selection of signal states $\mu_{a}$ ($\mu_{b}$) is a balance among quantum bit error rate, the probability of single-photon transmission, and error correction.
 

\begin{figure}[!htp]
    \centering
    \begin{minipage}[b]{0.49\textwidth}
        \centering
        \includegraphics[width=\linewidth]{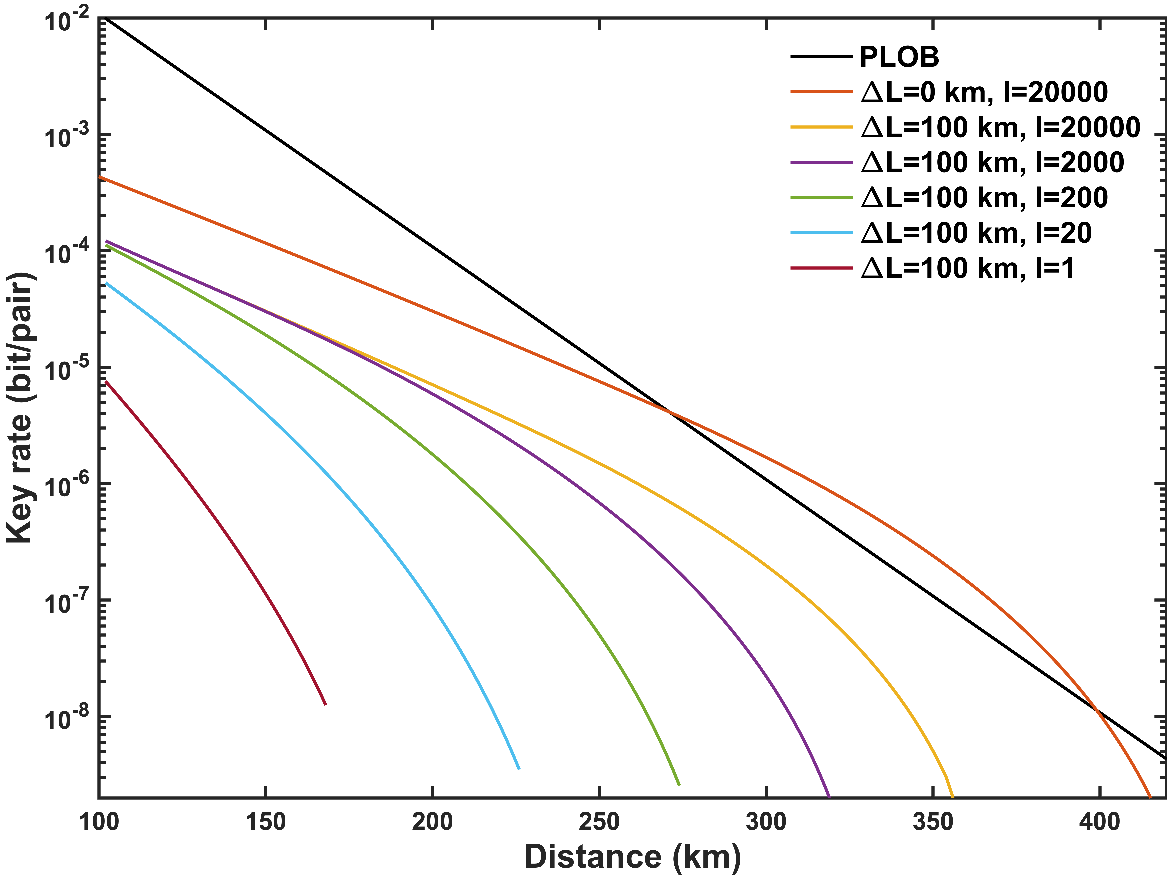}
        \caption{Optimized SKR with asymmetric intensity strategy versus the total distance ($L_A + L_B$) between Alice and Bob, at different maximal pairing intervals $l$.}
        \label{pairlen}
    \end{minipage}
    \hfill
    \begin{minipage}[b]{0.49\textwidth}
        \centering
        \includegraphics[width=\linewidth]{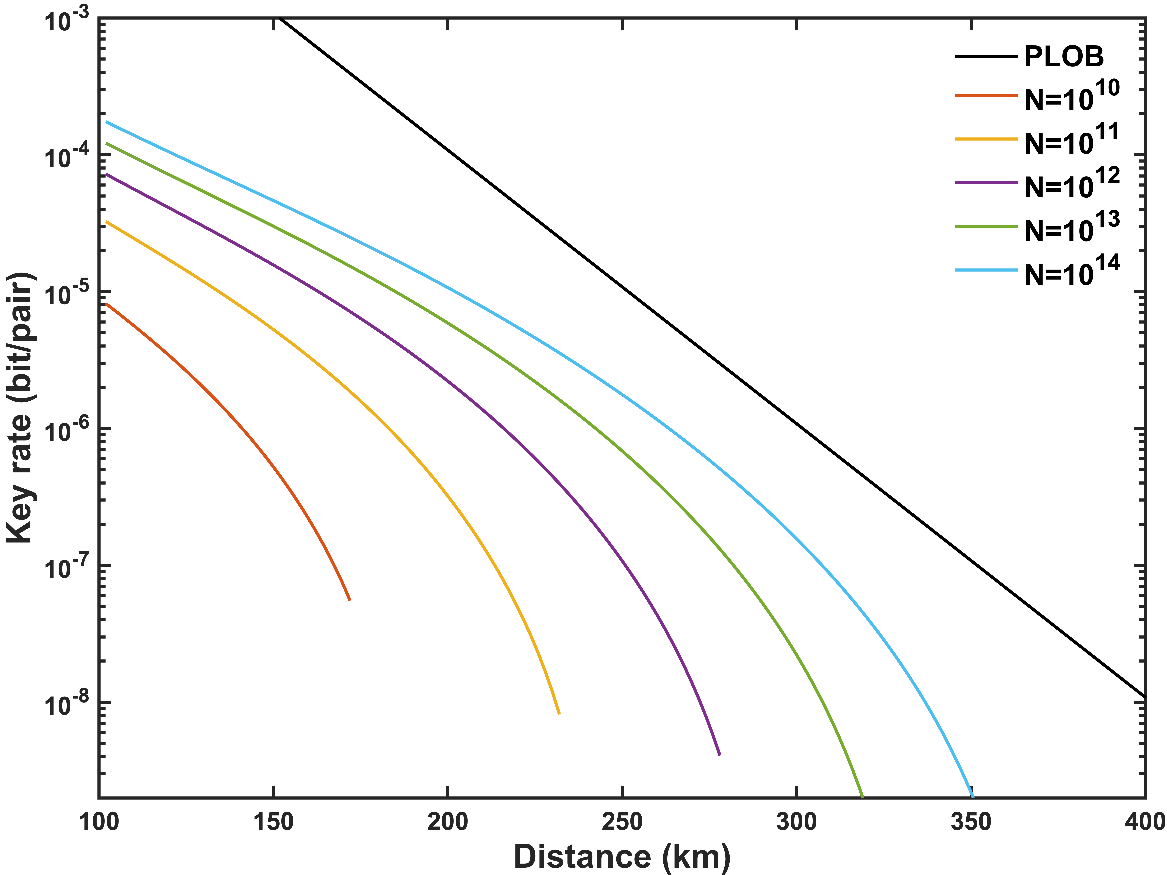}
        \caption{Optimized SKR with asymmetric intensity strategy versus the total distance ($L_A + L_B$) with different numbers of total pulses $N$ and $l=2000$.}
        \label{finitee}
    \end{minipage}
\end{figure}

 Figure \ref{probb} illustrates the variations in the probabilities of the signal, decoy, and vacuum states with asymmetric intensity strategy as the distance increases, when $\Delta L= 100$ km. It can be observed that as the distance increases, the probability of the signal states remains relatively stable in most cases, while the probability of the decoy states notably increases. As the distance increases and losses grow, an inevitable increase in the probability of the decoy states is necessary to achieve better HOM interference effects and lower error rates. 
 We also notice a relatively high proportion of vacuum states, attributable to their contribution in the pairing of the $Z$-basis, estimation of single-photon counts, and calculation of phase error rates.

Figure \ref{pairlen} demonstrates the influence of different maximum pairing intervals $l$ on the SKR with asymmetric intensity strategy. With the increase in $l$, the SKR in the asymmetric scenario shows a significant improvement but remains below that of the symmetric case, unable to surpass the PLOB bound. Additionally, as shown in Fig. \ref{finitee}, increasing the numbers of total pulses $N$ does not easily surpass the PLOB bound. While increasing $l$ significantly enhances the SKR in MP-QKD, the value of $l$ is constrained by the laser coherence time. 



\section{Conclusion}
In this paper, we demonstrate the performance of the asymmetric MP-QKD protocol with finite-key analysis. By combining universal composability security analysis and practical decoy state methods, our work aligns more closely with real-world experimental conditions. In the PSO process for asymmetric MP-QKD, deploying an asymmetric intensity strategy results in a noticeable improvement in the SKR compared to adding extra attenuation and is more practical for real-world deployment. However, the SKR still remains lower than in the symmetric case. Simulation results indicate that the decoy state $\nu_{a}$ ($\nu_{b}$) is more susceptible to HOM interference, closely aligning with ${\eta _a}{v_a} \approx {\eta _b}{v_b}$. However, the signal state $\mu_{a}$ ($\mu_{b}$), influenced by multiple factors, significantly deviates from this assumption. Increasing the maximum pairing interval $l$ can enhance the SKR in asymmetric MP-QKD, but it still falls significantly short compared to the symmetric case and remains unable to surpass the PLOB bound. 
Our research advances MP-QKD towards more practical network configurations and provides theoretical support for future MP-QKD quantum communication networks.

\begin{acknowledgments}
This work was supported by the Innovation Program for Quantum Science and Technology (2021ZD0301300); the State Key Laboratory of Information Photonics and Optical Communications (No. IPOC2024ZT10).
\end{acknowledgments}
\appendix

\section{The modified PSO algorithm for Maximizing $R(\vec{g})$}\label{mpso}
The PSO algorithm is a globally efficient and easily implementable optimization algorithm that reduces reliance on initial points through collaborative swarm behavior. It exhibits rapid convergence and is suitable for addressing a variety of complex multi-variable optimization problems. Considering the characteristics of the asymmetric MP-QKD protocol, we have modified the PSO algorithm as described in Algorithm \ref{alg2}. Lines 1-11 initialize the position, velocity, individual best, and global best of the particle swarm. It is reasonable and intuitive to expect that the particles, after initialization, should satisfy Eq. (\ref{condition1}). An often overlooked detail is that the intermediate parameters must also adhere to their physical significance as shown below
\begin{equation}\label{condition2}
\begin{aligned}
    &0 < n_{\left( k_a, k_b \right)}^Z, \ n_{\left( k_a, k_b \right)}^X, \ t_{\left( k_a, k_b \right)}^Z, \ t_{\left( k_a, k_b \right)}^X\\
       &0< y_{11}^{Z}, \frac{M_{\left( {\mu ,\mu } \right)}}{N}<1,\\
        &0<e_{11}^{Z,\rm{ph}}, E_{\left( {\mu ,\mu } \right)}<0.5.\\
\end{aligned}
\end{equation}
Equation (\ref{condition2}), which guarantees that all counts are positive, all count rates fall between 0 and 1, and all error rates are greater than 0 but less than 0.5, aligns with the requirements for practical deployment.
Compared to the original PSO algorithm \cite{Kennedy2010}, the modified version updates the inertia weight $w$, individual learning factor $c_1$, and social learning factor $c_2$ throughout the optimization process (Line 11). These updates improve the balance between global and local search, enhance particle coordination, and increase both efficiency and convergence speed. Lines 13-17 divide the particles into two groups: one group updates its position and velocity randomly, while the other follows the particle swarm optimization process. This effectively prevents the algorithm from getting trapped in local optima and enhances its global search capability.
Lines 19-24 involve updating the individual best positions $P_{\text{best},i}$, and the global best position $G_{\text{best}}$. When the termination condition is met, the returned $G_{\text{best}}$ and $R(G_{\text{best}})$ represent the optimal decoy state settings and SKR. The modified PSO algorithm, independent of the specific form or gradient information of the SKR function $R$, demonstrates strong robustness.  Its distinct advantage lies in its ability to handle complex, nonlinear problems, especially in high-dimensional spaces. 

\begin{algorithm}[!htbp]
\caption{The modified PSO Algorithm for Maximizing $R(\vec{g})$ with 12-Dimensional Variables}
\label{alg2}
\SetKwInOut{Input}{Input}
\SetKwInOut{Output}{Output}
\SetKw{KwTo}{to}
\SetKwFor{ForEach}{for each}{do}{end for}
\SetKwFor{For}{for}{do}{end for}

\Input{Particle number $N_{\text{PSO}}$, max iterations $T$, inertia weight range $[w_{\text{init}}, w_{\text{final}}]$, individual learning factor range $[c_{1,\text{min}}, c_{1,\text{max}}]$, social learning factor range $[c_{2,\text{min}}, c_{2,\text{max}}]$, global exploration ratio $h$, fitness change threshold $\zeta$}
\Output{Global best position $G_{\text{best}}$ and corresponding fitness $R(G_{\text{best}})$}

\SetKwFunction{FMain}{The modified PSO}
\SetKwProg{Fn}{Function}{:}{}
\Fn{\FMain}{
    \SetKwComment{Comment}{/*}{*/}

\textbf{Initialize swarm:} \;
Randomly initialize position $\vec{g}_i$ and velocity $V_i$ for each particle\;
Ensure that the initial positions $\vec{g}_i$ satisfy the constraints in Eq. (\ref{condition1}) and (\ref{condition2})\;

\textbf{Initialize best positions:} \;
\ForEach{particle $i$ \KwTo $N_{\text{PSO}}$}{
    Compute fitness $R(\vec{g}_i)$ and set $P_{\text{best},i} \gets \vec{g}_i$\;
}
Set global best $G_{\text{best}} \gets \arg\max R(P_{\text{best},i})$\;

\textbf{Iterative optimization:} \;
\For{$t = 1$ \KwTo $T$}{
    Update inertia weight $w$ using:
    \[
    w \gets w_{\text{init}} - \frac{w_{\text{init}} - w_{\text{final}}}{T} \times t
    \]
    Update learning factors $c_1(t)$ and $c_2(t)$:
    \[
    c_1(t) \gets c_{1,\text{min}} + (c_{1,\text{max}} - c_{1,\text{min}}) \cdot \left(1 - \frac{t-1}{T}\right)
    \]
    \[
    c_2(t) \gets c_{2,\text{min}} + (c_{2,\text{max}} - c_{2,\text{min}}) \cdot \left(1 - \frac{t-1}{T}\right)
    \]

    \ForEach{particle $i$ \KwTo $N_{\text{PSO}}$}{
        \If{$i < N_{\text{PSO}} \times h$}{
            Randomly update position $\vec{g}_i$ and velocity $V_i$, ensuring constraints are met\;
        }
        \Else{
            \For{$d = 1$ \KwTo $12$}{
                Update velocity and position:
                $
                V_{i,d} \gets w \times V_{i,d} + c_1(t) \times \text{rand1()} \times (P_{\text{best},i,d} - \vec{g}_{i,d}) + c_2(t) \times \text{rand2()} \times (G_{\text{best},d} - \vec{g}_{i,d})
                $
                \[
                \vec{g}_{i,d} \gets \vec{g}_{i,d} + V_{i,d}
                \]
            }
        }
    }

    \textbf{Update individual and global best:} \;
    \ForEach{particle $i$ \KwTo $N_{\text{PSO}}$}{
        Compute fitness $R(\vec{g}_i)$\;
        \If{$R(\vec{g}_i) > R(P_{\text{best},i})$}{
            Update $P_{\text{best},i} \gets \vec{g}_i$\;
        }
        \If{$R(\vec{g}_i) > R(G_{\text{best}})$}{
            Update $G_{\text{best}} \gets \vec{g}_i$\;
        }
    }

    \If{$|R(G_{\text{best}}) - R_{\text{new}}(G_{\text{best}})| < \zeta$}{
        Terminate\;
    }
}

\textbf{Output:} $G_{\text{best}}$ and $R(G_{\text{best}})$\;
}
\end{algorithm}

\section{Simulation formulas for asymmetric MP-QKD}\label{siumlation}

The average response probability during each round $p$ can be formulated as
\begin{equation}
    p= \sum\limits_{{k_{a}^{i}}{k_{b}^{i}}} {{p_{{k_{a}^{i}}}}{p_{{k^{i}_{b}}}}{q_{{k_{a}^{i}}{k_{b}^{i}}}}} ,
\end{equation}
where ${q_{{k_{a}^{i}}{k_{b}^{i}}}} = 2y\left[ {{I_0}\left( x \right) - y} \right]$, $y = \left( {1 - {p_d}} \right){e^{ - \frac{{k_a^i{\eta _a} + k_b^i{\eta _b}}}{2}}}$ and $x = \sqrt {{\eta _a}{k_{a}^{i}}{\eta _b}{k_{b}^{i}}}$. The expected pair number generated during each round can be expressed as
\begin{equation}
    {r_p} = \left[ {\frac{1}{{p\left[ {1 - {{\left( {1 - p} \right)}^l}} \right]}} + \frac{1}{p}} \right].
\end{equation}

For the $Z$-pair, where $\left( {{k_a},{k_b}} \right) \in \left\{ {\left( {\mu_{a} ,\mu_{b} } \right)} \right.$, $\left( {\mu_{a} ,\nu_{b} } \right)$, $\left( {\mu_{a} ,o_{b}} \right)$, $\left( {\nu_{a} ,\mu_{b} } \right)$, $\left( {o_{a},\mu_{b} } \right)$, $\left( {\nu_{a} ,\nu_{b} } \right)$, $\left( {\nu_{a} ,o_{b}} \right)$, $\left( {o_{a},\nu_{b} } \right)$, $\left. {\left( {o_{a},o_{b}} \right)} \right\}$, the count of effective detections can be expressed as
\begin{equation}
    n_{\left( {{k_a},{k_b}} \right)}^Z = \frac{{N{r_p}}}{{{p^2}}}\sum\limits_{\left( {{k_a},{k_b}} \right)} {{p_{k_a^i}}{p_{k_a^j}}{p_{k_b^i}}{p_{k_b^j}}{q_{k_a^ik_b^i}}{q_{k_a^jk_b^j}}}.
\end{equation}
For $\left( {{k_a},{k_b}} \right) \in \left\{ {\left( {\mu_a ,o_b} \right)} \right.$, $\left( {o_a,\mu_b } \right)$, $\left( {\nu_a ,o_b} \right)$, $\left( {o_a,\nu_b } \right)$, $\left. {\left( {o_a,o_b} \right)} \right\}$ in the $Z$-pair, the number of error-effective detections is denoted by $t_{\left( {{k_a},{k_b}} \right),0}^Z = \frac{{n_{\left( {{k_a},{k_b}} \right)}^Z}}{2}$, while for $\left( {{k_a},{k_b}} \right) \in \left\{ {\left( {\mu_a ,\mu_b } \right)} \right.,\left( {\mu_a ,\nu_b } \right),\left( {\nu_a ,\mu_b } \right),\left. {\left( {\nu_a ,\nu_b } \right)} \right\}$, it is denoted by 
\begin{widetext}
\begin{equation}
    t_{{\left( {{k_a},{k_b}} \right)},0}^Z{\rm{ }} = \frac{{N{r_p}}}{{{p^2}}}\sum\limits_{\left( {{k_a},{k_b}} \right),k_a^i = k_b^i = 0} {{p_{k_a^i}}{p_{k_b^i}}{p_{k_a^j}}{p_{k_b^j}}{q_{k_a^ik_b^i}}{q_{k_a^jk_b^j}}}  + \frac{{N{r_p}}}{{{p^2}}}\sum\limits_{\left( {{k_a},{k_b}} \right), k_a^j = k_b^j = 0} {{p_{k_a^i}}{p_{k_b^i}}{p_{k_a^j}}{p_{k_b^j}}{q_{k_a^ik_b^i}}{q_{k_a^jk_b^j}}} .
\end{equation}

For the $X$-pair, where $\left( {{k_a},{k_b}} \right) \in \left\{ {\left( {2{\mu _a},{o_b}} \right)} \right.$, $\left( {o_a,2\mu_b } \right),\left( {2\nu_a ,o_b} \right)$, $\left. {\left( {o_a,2\nu_b } \right)} \right\}$, the count of effective detections can be expressed as
\begin{equation}
    n_{\left( {{k_a},{k_b}} \right)}^X = \frac{{N{r_p}}}{{{p^2}}}\sum\limits_{\left( {{k_a},{k_b}} \right)} {{p_{k_a^i}}{p_{k_a^j}}{p_{k_b^i}}{p_{k_b^j}}{q_{k_a^ik_b^i}}{q_{k_a^jk_b^j}}}.
\end{equation}
The number of the error effective detections in this part can be represented as $t^X_{{\left( {{k_a},{k_b}} \right)},0} =\frac{{n_{\left( {{k_a},{k_b}} \right)}^X}}{2}$.

For $\left( {{k_a},{k_b}} \right) \in \left\{ {\left( {2\mu_a ,2\mu_b } \right)} \right.$, $\left( {2\mu_a ,2\nu_b } \right)$, $\left( {2\nu_a ,2\mu_b } \right)$, $\left. {\left( {2\nu_a ,2\nu_b } \right)} \right\}$ in the $X$-pair, the count of
effective detections can be expressed as
\begin{equation}
     n_{\left( {{k_a},{k_b}} \right)}^X \approx\frac{{N{r_p}}}{{{p^2}}}\frac{{2\Delta }}{\pi }{p_{k_a^i}}{p_{k_a^j}}{p_{k_b^i}}{p_{k_b^j}}\left( {4{y^4} - 8{y^3}{I_0}\left( x \right) + 2{y^2}\left( {{I_0}\left( {x\sqrt {2 - 2\cos \Delta } } \right) + {I_0}\left( {x\sqrt {2 + 2\cos \Delta } } \right)} \right)} \right),
\end{equation}
where ${I_0}\left(  \cdot  \right)$ is the modified Bessel function of the first kind. The error effective detections corresponding to this section can be expressed as
\begin{equation}
     t^X_{{\left( {{k_a},{k_b}} \right)},0} \approx \frac{{N{r_p}}}{{{p^2}}}\frac{{2\Delta }}{\pi }{p_{k_a^i}}{p_{k_a^j}}{p_{k_b^i}}{p_{k_b^j}}\left( {2{y^4} - 4{y^3}{I_0}\left( x \right) + 2{y^2}{I_0}\left( {x\sqrt {2 - 2\cos \Delta } } \right)} \right).
\end{equation}
Considering the misalignment-error, the error effective detections of the $Z$-pair and $X$-pair can be rewritten as
\begin{equation}
    \begin{aligned}
        t_{\left( {{k_a},{k_b}} \right)}^Z &= \left( {1 - e_d^Z} \right)t_{{\left( {{k_a},{k_b}} \right)},0}^Z + e_d^Z\left( {n_{\left( {{k_a},{k_b}} \right)}^Z - t_{{\left( {{k_a},{k_b}} \right)},0}^Z} \right),\\
        t_{\left( {{k_a},{k_b}} \right)}^X &= \left( {1 - e_d^X} \right)t_{{\left( {{k_a},{k_b}} \right)},0}^X + e_d^X\left( {n_{\left( {{k_a},{k_b}} \right)}^X - t_{{\left( {{k_a},{k_b}} \right)},0}^X} \right),\\
    \end{aligned}
\end{equation}
where ${e_d^Z}$ and ${e_d^X}$ is the misalignment-error of the $Z$-pair and $X$-pair, respectively.
\end{widetext}


\bibliography{apssamp}

\end{document}